\documentclass[aps,pra,twocolumn,superscriptaddress]{revtex4-1}

\usepackage{graphicx}
\usepackage{hyperref}
\usepackage{amsmath}
\usepackage{amssymb}
\usepackage{epstopdf}
\usepackage{xcolor}

\begin{document}

\title{Mid-Infrared Single-Photon Edge Enhanced Imaging\\ based on Nonlinear Vortex Filtering}

\author{Yinqi Wang}
\affiliation{State Key Laboratory of Precision Spectroscopy, East China Normal University, Shanghai 200062, China}

\author{Jianan Fang}
\affiliation{State Key Laboratory of Precision Spectroscopy, East China Normal University, Shanghai 200062, China}

\author{Tingting Zheng}
\affiliation{State Key Laboratory of Precision Spectroscopy, East China Normal University, Shanghai 200062, China}

\author{Yan Liang}
\affiliation{School of Optical Electrical and Computer Engineering, University of Shanghai for Science and Technology, Shanghai 200093, China}

\author{Qiang Hao}
\affiliation{School of Optical Electrical and Computer Engineering, University of Shanghai for Science and Technology, Shanghai 200093, China}

\author{E Wu}
\affiliation{State Key Laboratory of Precision Spectroscopy, East China Normal University, Shanghai 200062, China}
\affiliation{Chongqing Institute of East China Normal University, Chongqing 401121, China}

\author{Ming Yan}
\affiliation{State Key Laboratory of Precision Spectroscopy, East China Normal University, Shanghai 200062, China}
\affiliation{Chongqing Institute of East China Normal University, Chongqing 401121, China}

\author{Kun Huang}
\email{khuang@lps.ecnu.edu.cn}
\affiliation{State Key Laboratory of Precision Spectroscopy, East China Normal University, Shanghai 200062, China}
\affiliation{Chongqing Institute of East China Normal University, Chongqing 401121, China}

\author{Heping Zeng}
\email{hpzeng@phy.ecnu.edu.cn}
\affiliation{State Key Laboratory of Precision Spectroscopy, East China Normal University, Shanghai 200062, China}
\affiliation{Chongqing Institute of East China Normal University, Chongqing 401121, China}
\affiliation{Jinan Institute of Quantum Technology, Jinan, Shandong 250101, China}
\affiliation{CAS Center for Excellence in Ultra-intense Laser Science, Shanghai 201800, China}
\affiliation{Shanghai Research Center for Quantum Sciences, Shanghai 201315, China}

\begin{abstract}
Edge enhanced imaging via the spiral phase contrast enables to reveal the phase or amplitude gradients of a target, which has been proved useful in feature recognition, machine vision, and object identification. A long quest is to extend the operation wavelength into the mid-infrared (MIR) region, as highly demanded in various fields including infrared sensing, astronomic observation, and biomedical diagnosis. Here, we demonstrated ultra-sensitive MIR imaging at the single-photon level based on nonlinear frequency upconversion, where the spectrally converted replica of the MIR object image at 3070 nm was captured by a silicon electron multiplying charged coupled device. The imaging sensitivity was significantly improved by the coincidence pulsed pumping with a spectro-temporal optimization. Furthermore, the edge enhancement has been realized by imprinting the spiral phase pattern of the pump onto the upconverted field at the Fourier plane within the nonlinear crystal. Such a nonlinear spatial filter not only provided an effective way to implement the required high-fidelity vortex screening in the edge enhanced detection, but also rendered the MIR illumination into a visible image in an efficient and low-noise fashion. The presented system for MIR edge enhanced imaging might facilitate immediate applications in label-free histopathological diagnosis and non-destructive defect inspection.
\end{abstract}

\maketitle

\section{Introduction}

The spiral phase contrast (SPC) imaging constitutes an essential tool in modern image processing technology, which is commonly used in high-contrast microscopy \cite{Zernike1942Physica, Furhapter2005OE} and optical vortex coronagraph \cite{Foo2005OL}. Comparing to standard bight-field imaging systems, the SPC technique employs a vortex phase filter at the Fourier plane within a 4f optical relay configuration \cite{Khonina1992JMO}. The introduced phase hologram imprints a helical phase term of the form $e^{i\phi}$ on the spatial frequency components of the diffracted light field. Consequently, the resulted image exhibits a strong and isotropic edge contrast enhancement for both amplitude and phase objects \cite{Davis2000OL}. Although the edge enhanced operation could be realized by posterior calculation based on numerical spatial differentiation \cite{Zhu2017NC} or a two-dimensional radial Hilbert transform \cite{Guo2006OL}, yet the all-optical engineering of the point spread function provides a simpler and faster way to implement real-time imaging processing, thus favoring applications in machine vision, feature detection, and intelligent identification. Moreover, advanced functionalities, such as orientation-selective edge enhancement, are readily available by applying a symmetry-breaking central phase shifter for the zero-order Fourier component \cite{Jesacher2005PRL, Situ2009JOSAA}. The resulting shadow effect could provide a relief-like view of the sample topography with a sub-wavelength longitudinal resolution \cite{Jesacher2005PRL}. 

Typically, the SPC imaging is operated at the visible or near-infrared spectral bands due to the accessibility to high-performance phase modulators and imaging cameras. Nowadays, there is a significant impulse to extend the operation wavelength into the mid-infrared (MIR) region, pertaining to important applications including infrared monitoring, planetary observation, spectroscopic imaging, and biomedical diagnosis \cite{Fang2020LPR}. In this context, it is thus required to develop enabling techniques to approach sensitive MIR imaging camera and high-fidelity vortex filter. In general, the conventional MIR imagers based on indium antimonide (InSb) and mercury cadmium telluride (MCT) are high-cost and limited by the intrinsic noise, which often require cryogenic operation to suppress the intrinsic blackbody radiation and dark current. Particularly, superconducting nanowire has been used to demonstrate broadband MIR single-photon detection \cite{Marsili2012NL, Chen2020Arxiv}, albeit without demonstrating imaging capability. In parallel, a considerable effort has been dedicated to realizing room-temperature MIR detection, for instance resorting to emerging materials and novel structures based on black phosphorus \cite{Long2019SA, Bullock2018NP}, graphene plasmons \cite{Guo2018NM}, and quasi-2D tellurium \cite{TongNC2020}. However, it is still very challenging to approach the single-photon sensitivity with orders of magnitudes lower than the currently attainable noise equivalent power about pW/Hz$^{1/2}$.

An alternative promising solution lies in the nonlinear frequency upconversion, where the MIR signal is spectrally converted to the visible regime and subsequently detected by high-performance silicon sensors \cite{Rodrigo2021LPR, HuangPR2021}. A similar spirit was also manifested in MIR imaging based on non-degenerate two-photon absorption in wide-bandgap semiconductors \cite{Fishman2011NP, Knez2020LSA} or nonlinear interferometry with correlated dual-color photons \cite{Paterova2020SA, Kviatkovsky2020SA, Basset2021LPR}. Indeed, current advances in charged coupled devices(CCDs) have led to the development of electron multiplying CCDs (EMCCDs) with single-photon detection sensitivity and nearly-unitary quantum efficiency \cite{Zhou2013APL}. In previous works, the upconversion technique has enabled to demonstrate ultra-sensitive MIR imaging \cite{Zhou2013APL, Dam2012NP}, as well as direct extensions to video-rate hyperspectral imaging \cite{Junaid2019Optica} and high-resolution optical coherence tomography \cite{Israelsen2019LSA}. In addition to the efficient and sensitive MIR detection, the involved nonlinear conversion process also allows us to manipulate the incident field by engineering the pump properties in the spectral \cite{Fisher2016NC}, temporal \cite{Ansari2018Optica, Manurkar2016Optica}, and spatial domains \cite{Zhou2016LSA}. This unique feature has recently been exploited to implement nonlinear spatial filtering in SPC imaging, where the required spiral phase pattern could be transferred from the pump to the upconverted field at the Fourier plane within the nonlinear crystal \cite{Qiu2018Optica}. The underlying upconversion imaging processing was further used to realize edge enhanced detection \cite{Liu2019LP, Liu2019PRAp, Junaid2020AO} and nonlinear vortex coronagraph \cite{Engay2020OL} in the near infrared. In principe, the upconversion imaging technique could access to a longer operation wavelength \cite{Rodrigo2021LPR} and a broadband acceptance window \cite{Mrejen2020LPR}, thus circumventing the need for the MIR achromatic vortex mask that is currently hard to be fabricated \cite{Engay2020OL}.

In this paper, we have implemented, for the first time to the best of our knowledge, an ultra-sensitive MIR edge enhanced imaging based on the sum frequency generation. The involved nonlinear conversion facilitated spectral translation of incident MIR field into the visible region, where a high-performance silicon electron multiplying CCD was used for efficient and sensitive registration of upconverted photons. The imaging sensitivity at the single-photon level was made possible by the combination of improved conversion efficiency and reduced background noise in our coincidence-pumping configuration. More importantly, the vortex-carrying pump field enabled the high-fidelity mapping of spiral-phase mask onto the Fourier components during the three-wave mixing process. The demonstrated nonlinear spatial filter could provide an all-optical solution towards MIR achromatic operation for the spiral phase imaging, which would eliminate the stringent requirement of broadband vortex masks or hologram gratings. Therefore, the implemented MIR edge enhanced imaging system with the single-photon sensitivity might promote a variety of low-light-level applications in defect inspection, astronomical observation, biomedical examination, and defense surveillance.\\

\begin{figure}[b!]
\includegraphics[width=1\columnwidth]{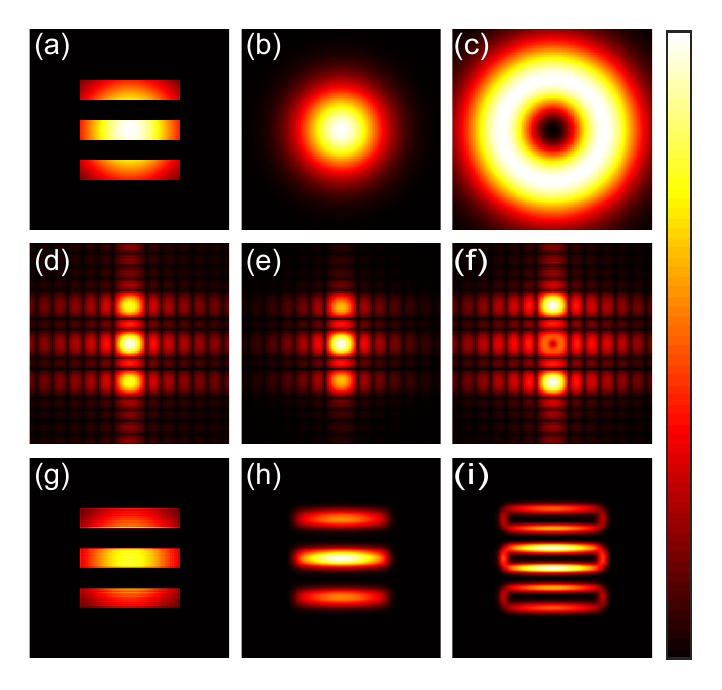}
\caption{Image simulation of the edge enhancement based on the nonlinear vortex filtering. Without the filtering at the Fourier plane of a 4f imaging system, the illuminated object (a) is focused to form the Fourier-transformed pattern (d), which is then inversely transformed back to the image (g). The presence of Gaussian beam (b) and vortex beam (c) at the Fourier plane leads to spatially filtered distributions as shown in (e) and (f), respectively. The corresponding reconstructed images are given in (h) and (i). The size for figures (d-f) is 1 mm, while that for the others  is 4 mm. Note that the square-root operation was applied on the field amplitude to enhance the image contrast in (d-f).}
\label{fig1}
\end{figure}

\section{Principles}
The core of the edge enhancement process lies in the manipulation of the spatial frequency components within a 4f imaging system. In the generic SPC imaging configuration, a vortex phase plate (VPP) should be placed at the Fourier plane to realize the phase modulation, which equivalently implements the two-dimensional radial Hilbert transformation filter \cite{Davis2000OL}. Since the VPP filter is typically operated at the focused plane, the required fabrication precision should be kept high at the tightly confined spatial region. Additionally, the VPP is usually realized by the azimuthal variation of optical thickness, or the vectorial polarization engineering based on crystal birefringence. The inevitable material dispersion requires the VPP to operate at a narrow wavelength band for approaching a high-fidelity vortex phase pattern. Indeed, broadband achromatic VPPs remains challenging to be fabricated, especially in the MIR regime \cite{Engay2020OL}. 

\begin{figure*}[t!]
\includegraphics[width=0.84\textwidth]{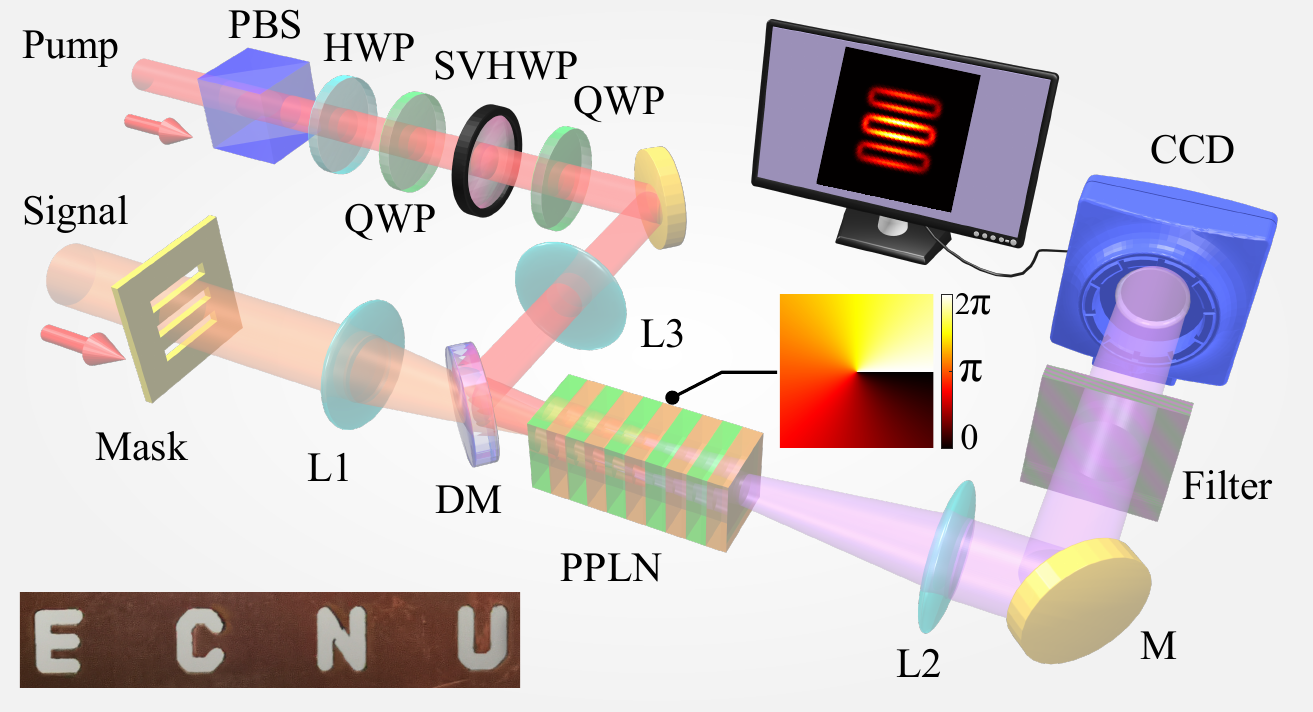}
\caption{Experimental scheme for the MIR nonlinear edge enhanced imaging. The object image formed by the collimated MIR illumination propagates through a 4f configuration based on two lenses (L1 and L2). Comparing to the conventional spiral phase imaging, the required vortex phase filter here is prepared by a pump source carrying an orbital angular momentum. The resultant modulation pattern at the Fourier plane is featured with a doughnut-shape intensity and a helical phase distribution. The implemented nonlinear phase filtering not only provides high-fidelity engineering of the Fourier frequency components to realize edge enhancement, but also leads to efficient wavelength conversion into the visible band for subsequent sensitive detection. The image at the bottom-left corner shows the hollowed-out copper sheet with an acronym for the university. HWP: half-wave plate; QWP: quarter-wave plate; SVHWP: spatially varied HWP; PBS: polarization beam splitter; DM: dichroic mirror; M: silver mirror; PPLN: periodically poled lithium niobate crystal.}
\label{fig2}
\end{figure*}

In this context, here we focus on an alternative all-optical method based on sum frequency generation (SFG). Under the paraxial and slowly varying envelope approximation, the SFG field $E_\text{up}$ could be deduced from the wave coupling equation as
\begin{equation}
 \frac{\partial  E_\text{up}} { \partial z} =  \frac{i 2 \omega^2_\text{up} d_\text{eff}}{k_\text{up} c^2} \times E_s E_p e^{-i \Delta k_z z} \ ,
\end{equation}
where $E_s$ and $E_p$ are the signal and pump electric fields, $d_\text{eff}$ denotes the effective nonlinear coefficient, $c$ is the speed of light in vacuum, $z$ is the propagation direction, and $\Delta k_z$ represents the longitudinal phase mismatch of the nonlinear conversion process. The involved angular frequencies for the signal, pump and SFG fields follow the energy conservation as $\omega_\text{up} = \omega_\text{s} + \omega_\text{p}$. Experimentally, the spectral bandwidth of the signal could be optimized within the phase-matching bandwidth to improve the conversion efficiency. The resulting expression for the SFG field would be simply related to the product of the signal and pump fields. Consequently, the nonlinear wave mixing not only results in the wavelength conversion, but also enables to map the phase from the pump to upconverted field \cite{Zhou2016LSA}. The all-optical phase manipulation lays the foundation to implement the so-called nonlinear vortex filter \cite{Qiu2018Optica}.

More specifically, we consider a signal object image described by $E_s(r, \phi)$, which is Fourier transformed by the first lens in the 4f image system to be $\tilde{E}_s(\rho, \varphi) = \mathcal{F}[E_s(r, \phi)]$. Then a pump field $E_p(\rho, \varphi)$ at the focal plane was applied as an effective filter to the spatial frequency components, resulting in a total field given by
\begin{equation}
\tilde{E}_f(\rho, \varphi) = \tilde{E}_s(\rho, \varphi) \times E_p(\rho, \varphi) \ .
\end{equation}
After the second lens, the inverse Fourier transform can be performed to obtain the field at the image plane as
\begin{equation}
E_\text{up}(r, \phi) = E_s(r, \phi) \otimes \mathcal{F}[E_p(\rho, \varphi)] \ ,
\end{equation}
where $\otimes$ denotes a two-dimensional convolution. It can seen that the Fourier transform of the pump field actually acts as the point spread function (PSF). In the case of a Gaussian pump shown in Fig. \ref{fig1}(b), the PSF also follows the Gaussian distribution. The upconverted field can simply be given by \cite{Dam2012NP}
\begin{equation}
E_\text{up}(x,y) \propto E_s(-\frac{x}{M} , -\frac{y}{M}) \otimes \exp{[- \frac{\pi w_0^2 (x^2+y^2)}{\lambda^2_{up} f_2^2}]} \ ,
\label{eq4}
\end{equation}
where $M =  \lambda_{up} f_2 / (\lambda_{s} f_1)$ is the magnification factor, $f_1$ and $f_2$ are the focal lengths of the first and second lenses in the 4f system, $w_0$ is the pump beam radius. As indicated in Fig. \ref{fig1}(e), the limited-size pump serves as a soft aperture, which leads to the image blurring in Fig. \ref{fig1}(h) due to the attenuation of the high-frequency components.

\begin{figure*}[t!]
\includegraphics[width=0.95\textwidth]{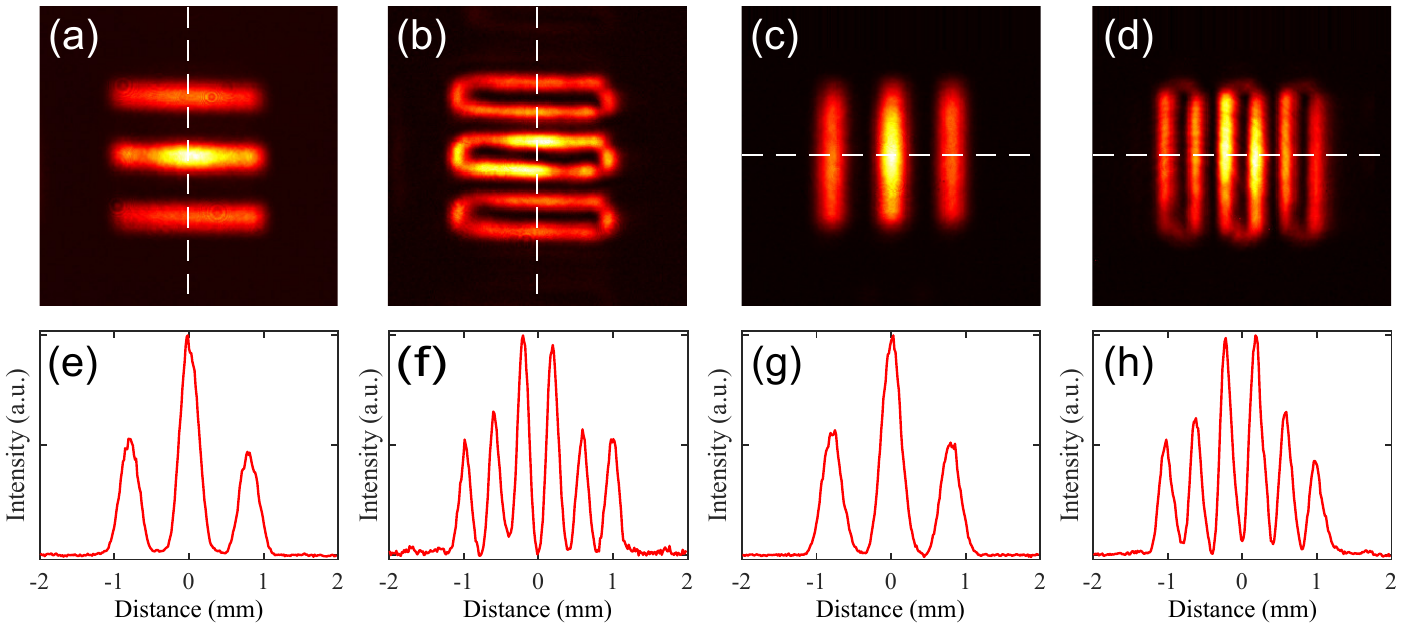}
\caption{Experimental upconversion images of the horizontal and vertical lines in a USAF target. (a,c) present the typical bright-field images at the presence of Gaussian pump filtering, while (b,d) show the acquired images with the nonlinear vortex filtering. (e-h) give the corresponding the normalized intensity distribution along the particular direction. The implemented all-optical spiral phase contrast technique clearly reveal the edge enhancement effect.}
\label{fig3}
\end{figure*}

Similarly, if we consider a pump field with a helical phase distribution as $E_p(\rho, \varphi) = \text{circ}(\rho/R) \exp(i \varphi)$, where $\text{circ}(\rho/R)$ denotes the aperture function with a radius of $R$. As shown in Fig. \ref{fig1}(e), the presence of the on-axis phase singularity results in a null intensity at the center of the beam \cite{Liu2019PRAp}. Consequently, the low spatial frequency components are significantly dimmed as illustrated in Fig. \ref{fig1}(f). In this case, the PSF for the upconversion imaging can be written as
\begin{equation}
\text{PSF} = \frac{\pi R}{2 r} [J_0(\sigma r) H_1(\sigma r) - J_1(\sigma r) H_0(\sigma r)] e^{i \phi}   \ ,
\label{eq5}
\end{equation}
where $\sigma = 2 \pi R/ (\lambda_\text{up} f_2)$, $J_m$ is the $m^\text{th}$-order Bessel function of the first kind, and $H_m$ is the $m^\text{th}$-order Struve function \cite{Khonina1992JMO}. The resulting PSF is featured with a doughnut-shaped intensity ring and an azimuthally-varied phase pattern. Such a kernel function with an angular coordinate $\phi$ dependence guarantees that the destructive interference occurs for any two image points located at central-symmetric positions \cite{Furhapter2005OE}. Hence, the convolution integral produces intensity maxima at amplitude or phase edges of an image as shown in Fig. \ref{fig1}(i). Note that the edge enhancement effect is isotropic to all directions due to the spherical symmetry of the kernel. Intriguingly, off-center spatial filtering operation can be used to implement orientation-selective edge enhancement detection \cite{Situ2009JOSAA, Xu2020OE}. The above discussed effects have been investigated experimentally, as detailed in the following sections.

\section{Experimental Section}
Figure \ref{fig2} presents the artistic illustration of the experimental setup for the MIR edge enhancement imaging based on nonlinear frequency conversion. The involved light sources were originated from a passively synchronized fiber laser system \cite{HuangPR2021}, which consisted of an Yb-doped fiber laser (YDFL) at 1030 nm and an Er-doped fiber laser (EDFL) at 1550 nm. Two dual-color pulses were then used to prepare the MIR signal source at 3070 nm based on difference frequency generation (DFG).  Typically, it is challenging to measure weak MIR light due the limited powermeter sensitivity and ambient thermal disturbance. The intrinsic linear response of the coherent downconverter provided an effective way to precisely calibrate the MIR power, particularly at the low-light level (see the Supporting Information).

As for the pump source, the average power of the YDFL output was boosted to 700 mW by two-stage fiber amplifiers. Then a spiral phase distribution $e^{i m \varphi}$ of the complex pump field was prepared by using a spatially varying half-wave plate (SVHWP) with a unitary topology charge $m=1$. The underlying mechanism is related to the so-called Pancharatnam-Berry geometrical phases involved in the azimuthally variant transformation of the optical polarization \cite{Junaid2020AO}. As shown in Fig. \ref{fig2}, the SVHWP was sandwiched between two quarter-wave plates with the diagonally orientated optical axis, which would impose a vortex phase on the Gaussian beam while keeping the original polarization as detailed in the Supporting Information. The vertical polarization direction for the involved beams was required to fulfill the type-0 phase-matching condition. 

Finally, the signal and pump beams were spatially combined by a dichroic mirror before being steered into a periodically poled lithium niobate (PPLN) crystal with a length of 10 mm. The upconverted light at 771 nm passed through a series of spectral filters with a total transmission of 82\% and 218-dB rejection for the pump. In combination with the coincidence-pulse pumping, the parametric background noise was significantly within an ultrashort time window, which enabled the subsequent single-photon imaging. The spectro-temporal properties of relevant pulses were engineered to improve the nonlinear conversion efficiency \cite{HuangPR2021}. Specifically, a spectrally bright MIR source with a narrow band was tailored to approach the phase-matching bandwidth. And the pulse duration of the pump is optimized to be longer than that for the signal, thus ensuring the high intensity for the wrapped MIR photons. More information about the experimental setup and related results for the pulse characterization are given in the Supporting Information.

\section{Results}
Now we turn to characterize the frequency upconversion imaging system. As shown in Fig. \ref{fig2}, a clear optical path USAF-1951 resolution target was illuminated by the collimated MIR light. The formed object image was then focused ($f_1$ = 50 mm) at the center of the nonlinear crystal. The mixed pumped beam provided an all-optical filter in the 4f imaging system. After another lens ($f_2$ = 200 mm), the upconverted image was scaled by the factor $M =  \lambda_\text{up} f_2 / (\lambda_\text{s} f_1) \approx 1$, and captured by a low-noise and high-speed silicon CCD (Newport, LBP2-HR-VIS2). Figures \ref{fig3}(a) and (c) presents the captured images for the horizontal and vertical lines under a Gaussian pump. The three-bar pattern is the third element of group 0 in the test target and corresponds to a spatial frequency of 1.26 lp/mm. The blurring effect was observed due to the high-frequency attenuation as described by Eq. \ref{eq4}. The imaging resolution was about 140 $\mu$m, which was mainly limited by the beam size of the pump within the nonlinear crystal. A better resolution is possible by enlarging the waist of the focused pump beam, albeit that the conversion efficiency would be degraded due to the lower pump intensity \cite{Dam2012NP}. In the presence of vortex pumping, the image outlines were highlighted by the SPC-based process, as shown in Figs. \ref{fig3}(b) and (d). The resultant edge enhancement imaging was manifested clearly in the intensity distribution along the section line. These experiment results agreed well with the theoretical simulations in Fig. \ref{fig1}. Notably, both the edge enhancement and optimized resolution were related to the engineering of the Fourier components. The edge enhanced performance lied in the interferometric cancellation of the low spatial frequencies, while the better resolution was obtained by keeping the high-frequency components as much as possible. Therefore, the adopt of a higher vortex topology would lead to implement the so-called curved edge detection for sharp corners along with an enhanced resolution \cite{Liu2020OE}.

\begin{figure}[b!]
\includegraphics[width=1\columnwidth]{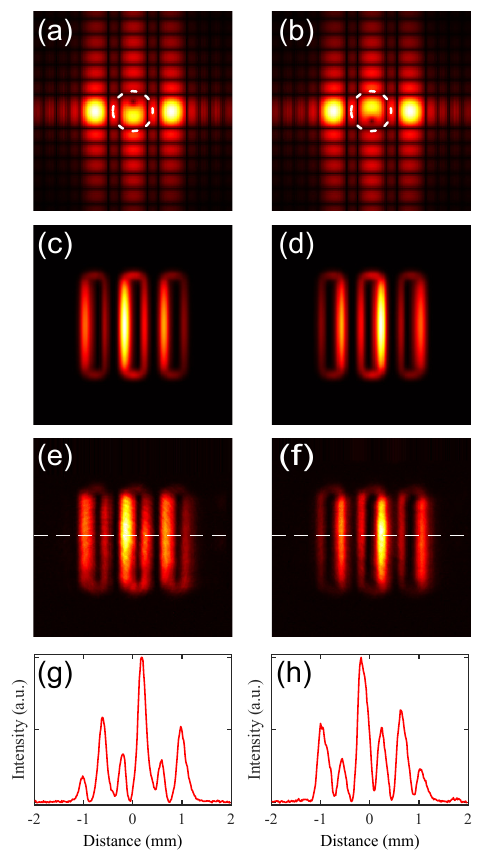}
\caption{Orientation-selective edge enhancement for MIR imaging. (a,b) show the filtered patterns at the Fourier plane in the 4f imaging system by up- and down-shift the vortex filter away from the center. (c,d) present the theoretical simulation for the off-axis filtering, while (e,f) give the corresponding experimentally acquired images. The intensity distribution along the cross section indicate the selective edge detection.}
\label{fig4}
\end{figure}

\begin{figure*}[t!]
\includegraphics[width=0.6\textwidth]{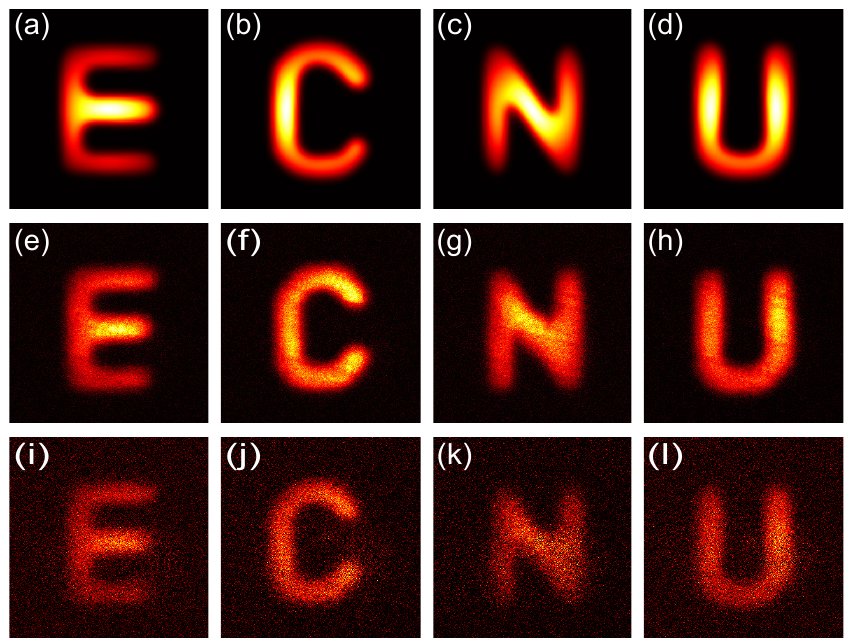}
\caption{MIR upconversion imaging at the single-photon level. (a-d) give the theoretical simulation for the illuminated mask with four letters. (e-h) show the experimentally acquired bright-field MIR images with an average photon number per pulse of 5. (i-l) present the imaging performance at the single-photon level. The incident MIR pulse contains an average photon number of 0.5. The figure size is 4 mm.}
\label{fig5}
\end{figure*}

Furthermore, we have investigated the orientation-selective edge enhancement by laterally shifting the pump vortex mask at the Fourier plane. The off-axis spatial filtering process allows to selectively manipulate the selective spatial frequencies \cite{Situ2009JOSAA}. In this scenario, the circular symmetry breaking of the spiral phase distribution would result in a so-called shadow effect, $\textit{i.e.}$, the edges of a sample object are highlighted depending on the local direction \cite{Jesacher2005PRL}. According to the simulation given in Figs. \ref{fig4}(a-d), the shadow orientation is perpendicular to the off-axis direction of the vortex filter. The corresponding experimental observations are given in Figs. \ref{fig4}(e,f), exhibiting enhanced or attenuated edges from one side to the other. This anisotropic filtering effect could also been identified from the uneven amplitudes for the two edges in a given line of the object as illustrated in Figs. \ref{fig4}(g,h), which was in contrast to the cross-section distribution given in Fig. \ref{fig3}(h).

Next, we turn to demonstrate the competence of the ultra-sensitive MIR imaging at the low light-level. A transmission mask with four characters ``E", ``C", ``N", and ``U" was used to prepare the object beams, as shown in the bottom-left corner of Fig. \ref{fig2}. Then the visible SFG photons were registered by a megapixel back-illuminated EMCCD (Andor, iXon Ultra 888) with a spatial resolution of 13 $\mu$m. The quantum detection efficiency was about 60\% at 771 nm. In the meanwhile, the dark noise was significantly suppressed by using a thermoelectric cooling down to -80 $^\circ$C. Even at the 1000-fold high gain operation, the dark current of the EMCCD was specified to be about 10$^{-4}$ electrons/pixel/second. Comparing to the MIR focal plane array based on HgCdTe or InSb, the used silicon camera thus makes it possible to realize an unequalled image performance in terms of high efficiency, low noise, fast speed, and high resolution \cite{Dam2012NP, Zhou2013APL}. Under the Gaussian pump, the acquired upconversion images were presented in Figs. \ref{fig5}(e-h), which were consistent to the theoretically simulated results in Figs. \ref{fig5}(a-d). The incident average photon number for the MIR signal pulse was set to be 5, and the accumulation time is 2 s. The presented experimental images were corrected for the background noises due to the intensive pump and ambient scattering photons. As shown in Fig. \ref{fig5}(i-l), further reduction of the photon number to 0.5 was still allowed to obtain high-contrast images at the condition of a longer accumulation time up to 7 s. In comparison to the near-infrared counterpart, the MIR upconversion detector is barely affected by the Raman scattering noise due to the large spectral separation between the pump and signal wavelengths. Also, the pedestal noise due to the crystal poling error is relatively smaller for the longer wavelength since the phase-matched gating period is larger. The dominant noise source is thus ascribed to the black body radiation from the heated nonlinear crystal, which would be more prominent for longer operation wavelengths \cite{Dam2012NP}.

\begin{figure*}[t!]
\includegraphics[width=0.6\textwidth]{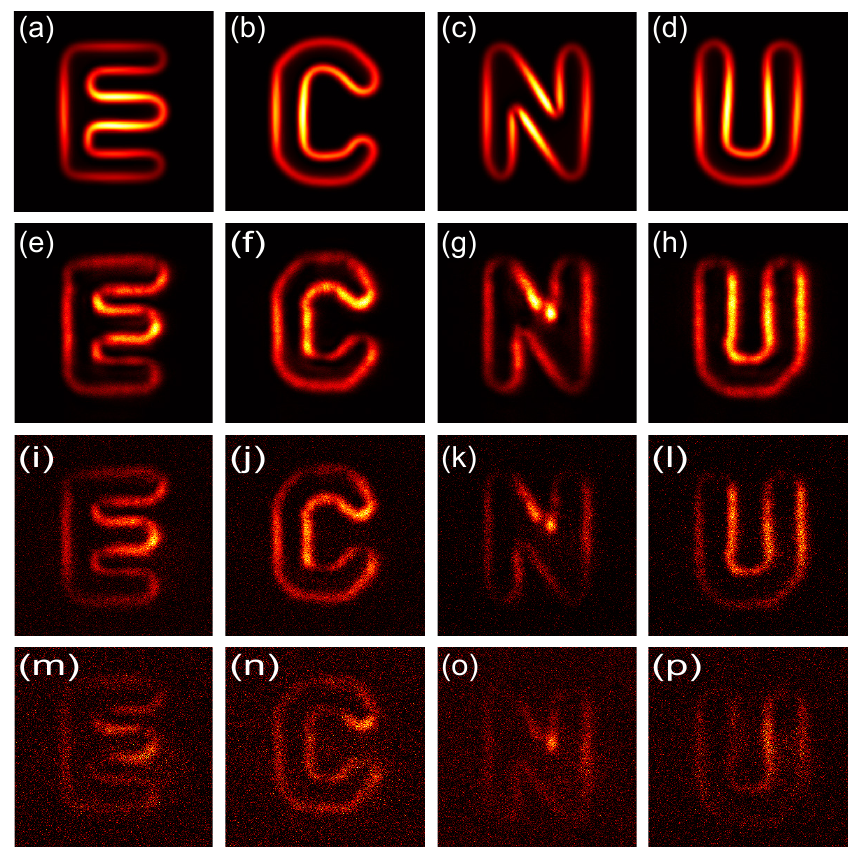}
\caption{Ultra-sensitive MIR edge-enhanced imaging based on nonlinear frequency conversion. Theoretical simulation (a-d) and experimental observation (e-h) are investigated for the illuminated object with four letters. The integration time of 20 ms is sufficient to acquire the high-contrast images with an incident pulse energy of 1 fJ. (i-l) and (m-p) present the low-light-level performance when the incident pulses contain 10 and 5 photons, respectively. The exposure time was set to 10 s. The size is 4 mm for all the figures.}
\label{fig6}
\end{figure*}

Finally, we have investigated the imaging performance for the edge enhancement by switching the pump to the vortex beam. Figures \ref{fig6}(a-d) give the simulated images for the four letters. More details on the numerical simulation and parameter setting are available in the Supporting Information. The corresponding experimental observations are presented in Figs. \ref{fig6}(e-h), clearly showing the hollow effect for the letters. The integration time was set to be 20 ms for an incident pulse energy of 1 fJ, thus allowing to achieve real-time and sensitive MIR imaging at the video frame rate. In the case of an extremely low illumination, it was still possible to reveal the target image, albeit that a longer exposure time was required to accumulate sufficient photons. Figures \ref{fig6}(i-l) present the acquired images in the presence of an input MIR pulse with an average photon number of 10. The photon number for the pulsed light flux onto each illuminated pixel was estimated to be smaller than 6$\times 10^{-4}$, which validated the single-photon operation. The superior sensitivity was further verified by using an input MIR illumination with 5 photons per pulse as shown in Figs. \ref{fig6}(m-p). The exhibiting noisy speckles became more pronounced due to the reduced signal-to-noise ratio. The limited image contrast could be ascribed to the decreased total detection efficiency. In the experiment, the average conversion efficiency for the four letters was measured to be about 5.1\% and 0.2\% under the Gaussian and vortex pumping, respectively. As described in the Supporting Information, the insertion of the SVHWP into the pump path would lead to an expanded intensity distribution within a doughnut shape. In the condition of constant average pump power, the peak intensity of the vortex beam will be reduced to 21\% comparing to that for the Gaussian pump. In addition, the null intensity of the vortex beam would block the low-frequency components at the Fourier plane, which resulted in 73\% energy loss during the filtering process. To go beyond the achieved result, higher pump power would be favorable with the trade-off optimization to the pump-induced noise \cite{Zhou2013APL}. The unprecedented sensitivity of MIR edge enhanced imaging achieved here would open up new possibilities to implement applications requiring operation at the photon-starving scenarios.

It is noteworthy that the superior sensitivity of the implemented upconversion imaging system benefited from the coincidence-pumping technique with ultrafast optical pulses. Similar mechanism was manifested in recent works on single-photon edge enhanced detection at 810 nm \cite{Liu2020OE, Zhou2020SA}, where the asynchronized background photons were significantly suppressed by the narrow time window for the two-photon coincidence. In comparison, the optical pulse could provide a much narrower gate for the imager than that given by the electrical pulse.

\section{Conclusion}
In conclusion, we have implemented an ultra-sensitive edge enhanced MIR imaging based on nonlinear frequency conversion. The involved nonlinear mixing between the MIR signal and the intensive pump field facilitated the simultaneous realization of the disparate wavelength transduction and the spiral phase mapping. Consequently, high-performance MIR imaging was demonstrated with the single-photon sensitivity, video-rate speed, and room-temperature operation. Moreover, the all-optical vortex filter was implemented by imprinting the helical phase pattern of the OAM-carrying pump onto the Fourier components, which was featured with high fidelity and achromatic operation. In combination with the adiabatic conversion technique \cite{Mrejen2020LPR}, high-fidelity vortex phase mask can be prepared in a broadband spectral range, thus avoiding the technically challenging fabrication to access the MIR achromatic filter \cite{Engay2020OL}. Additionally, extended field of view for the upconversion imaging was feasible by resorting to broadband pumping \cite{Junaid2020AO}, thermal gradient operation \cite{Liu2019PRAp} or angular rotation \cite{Junaid2019Optica} of the nonlinear crystal. The presented system for MIR single-photon edge enhanced imaging might simulate a variety of low-light-level applications, such as photon-starving night vision, long-distance infrared sensing, deep-tissue medical diagnosis, and phototoxicity-sensitive biological imaging.

\section*{Supporting Information}
Supporting Information is available from the Wiley Online Library or from the author.

\section*{Acknowledgements}
This work was supported by National Key Research and Development Program (2018YFB0407100), Science and Technology Innovation Program of Basic Science Foundation of Shanghai (18JC1412000),  National Natural Science Foundation of China (11621404, 11727812), Shanghai Municipal Science and Technology Major Project (2019SHZDZX01), Special Appointment (Eastern Scholar) at Shanghai Institutions of Higher Learning, and Fundamental Research Funds for the Central Universities.

\section*{Conflict of Interest}
The authors declare no conflict of interests.

\section*{Keywords}
edge enhancement, mid-infrared imaging, frequency upconversion, orbital angular momentum

\end{document}